\documentclass[twocolumn,superscriptaddress,prl]{revtex4}
\usepackage{amsfonts,amsmath,amssymb}
\usepackage{graphicx}

\begin{document}
\title{A Bounded Rational Driver Model}
\author{Ihor Lubashevsky}
 \affiliation{Theory Department, General Physics Institute, Russian Academy of
Sciences, Vavilov Str. 38, Moscow, 119991 Russia}
 \affiliation{Institute of
Transport Research, German Aerospace Center (DLR), Rutherfordstrasse 2, 12489
Berlin, Germany.}
\author{Peter Wagner}
\affiliation{Institute of Transport Research, German Aerospace Center (DLR),
Rutherfordstrasse 2, 12489 Berlin, Germany.}
\author{Reinhard Mahnke}
\affiliation{Fachbereich Physik, Universit\"at Rostock, D--18051 Rostock,
Germany}
\date{\today }

\begin{abstract}
  This paper introduces a car following model where the driving scheme
  takes into account the deficiencies of human decision making in a
  general way. Additionally, it improves certain shortcomings of most
  of the models currently in use: it is stochastic but has a
  continuous acceleration. This is achieved at the cost of formulating
  the model in terms of the time derivative of the acceleration,
  making it non-Newtonian.
\end{abstract}

\maketitle


To understand traffic flow, it is mandatory to analyze the interaction
between the cars. The simplest case is that of a car following a lead
car. To describe this process, a big number of models have been
invented (for a review see~\cite{Chowdhury,Helbing}). These models
differ in the details of the interaction between the cars, and the
time update rule, ranging from differential equations to cellular
automata. Mostly, they describe this process by an equation
$a=a\left( v,h,V \right)$ that relates the change in the current
velocity $v$ (the acceleration $a$) to the velocity $v$ of the
following car, the distance $h$ (``headway'') to the car ahead, and
its speed $V$, respectively.

\begin{figure}[tbp]
  \includegraphics[width=70mm]{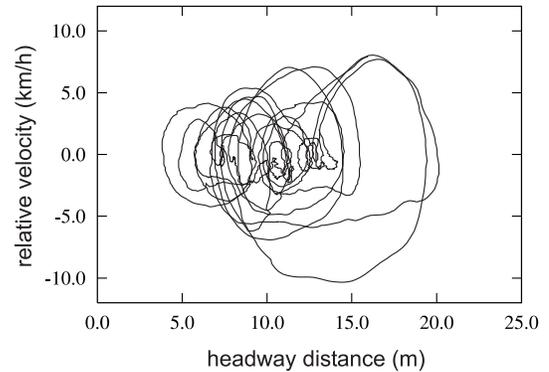}
\caption{Measured car-following behavior. Data are recorded by an
  equipped car measuring distance $h$ and speed $v$ and computing $v -
  V$ during a drive on a German freeway.
  \label{F0}}
\end{figure}

Considerable effort has been invested to investigate the emerging
macroscopic behavior from the underlying microscopic dynamics of
interacting cars.  Nevertheless, there is still a lot of controversy
in both the macroscopic behavior when compared to reality
\cite{daganzo-critique-sync}, and in the microscopic foundations of
the individual car dynamics. In particular, the observed non-damped
oscillations in the relative motion of vehicles, which are illustrated
in Fig.~\ref{F0}
are often explained by the instability in the cooperative motion of
the car ensemble only (see, e.g., \cite{Chowdhury,Helbing}). In fact,
subjected to reasonable physical constraints the relation $a=a\left(
v,h,V\right)$ seems to be hardly able to predict an instability in the
following car motion provided the car ahead moves at a constant
velocity. However, recent models \cite{BL,Tomer,Kerner} display a
certain kind of instability in the car following process itself.

There are actually two stimuli affecting the driver behavior. One of them is
the necessity to move at the mean speed of traffic flow, i.e., with the speed
$V$ of the leading car. So, first, the driver should control the velocity
difference $v-V$. The other is the necessity to maintain a safe headway
$h_{\text{opt}}(V)$ depending on the velocity $V$. In particular, the earliest
``follow-the-leader'' models \cite{83,84} take into account the former stimulus
only without regarding the headway $h$ at all. By contrast, the ``optimal
velocity'' model \cite{B1,B2} directly relates the acceleration $a$ to the
difference between the current velocity $v$ and a certain optimal value
$\vartheta _{\text{opt}}(h)$ at the current headway, $a\propto \lbrack
v-\vartheta _{\text{opt}}(h)]$. Of course, more sophisticated approximations,
e.g., \cite{Hel,Fritz,Xing,skPRE,H1,H2} to name but a few, allow for both
stimuli.

It is not very likely that the variables $\left\{ v,h,V\right\}$ do specify the
acceleration $a$ completely. Since drivers have motivations and follow only
partly physical regularities, memory effects may be essential. In a simple
manner, this has been introduced in models that relate the current acceleration
$a(t)$ to the velocity $v(t-\tau_a)$ and the headway $h(t-\tau_a)$ at a
previous moment $t-\tau_a$ (for a review of the ``following-the-leader'' models
see, e.g., Ref.~\cite{D1,D3}, for the ``optimal velocity model'' see
Ref.~\cite{D2}). Here, $\tau_a$ is the delay time in the driver response which
is treated as a constant. This approach is not completely satisfactorily, since
first, the physiological delay in the driver response seems to be too short to
be of importance. Second, it is not clear why the memory effects relate only
two moments of time instead of a certain interval as a whole. Third, the
dependence of the time scale $\tau_a$ on the car motion state is missing.
Nevertheless, these models show an instability in the car-following dynamics
(provided $\tau_a$ is big enough) and are non-Newtonian as well.

In the following, reasons of another nature than the driver response
delay lead beyond the framework of Newton's mechanics. A corresponding
model for the following car dynamics displaying an instability around
the stationary motion is proposed. To describe the driver behavior,
the approach suggested in Ref.~\cite{we1} will be used. There, drivers
plan their behavior for a certain time in advance instead of simply
reacting to the surrounding situation. A similar idea related to the
optimum design of a distance controlling driver assistance system is
discussed in Ref.~\cite{q}. In mathematical terms the driver's
planning of her further motion is reduced to finding extremals of a
certain priority functional that ranks outcomes of different driving
strategies. Here, the assumption that the driver is rational plays the
crucial role. It means that the driver continuously correct the car
motion to follow the optimal strategy. In this case \cite{we2}, the
collection of variables $\left\{ v,h,V\right\}$ does specify the car
acceleration $a$ completely. However, the assumed continuous control
is impossible to achieve for humans. Therefore, it is assumed below
that a real driver, first, cannot compute the optimal path of motion
exactly and, second, that she cannot correct the car motion
continuously.

This is just the approach that is known as bounded rationality
\cite{BR1}. Even if a driver succeeds in finding the optimal solution, 
she is only capable of
%
setting the acceleration to a fixed value. After that, she waits until
the deviation from her priority functional has become too big to
ignore, leading to a re-computation of another more or less optimal
path. Or, to put it differently, drivers are simply not capable of
resolving small differences between a given value of acceleration,
speed, or headway and their ``optimal'' desired values.

The re-computations are assumed to happen stochastically, with a
probability that increases with the deviation from the desired
state. So, the model described below becomes a stochastic one. The
action of noise can be modeled either explicitly by introducing
certain thresholds (as is done in the psycho-physical
models~\cite{Fritz}) or by making the noise amplitude dependent on the
distance between the current and optimal state. This defines a dynamic
trap model \cite{we3}, an approach that will be followed below.

To make the model more realistic, it is demanded that the trajectories
of acceleration, speed, and headway are continuous functions of time.
This can be achieved by formulating the model in terms of the
time-derivative of acceleration called jerk  
and adding a white-noise term there. In what follows, that the
acceleration is a colorized noise process without jumps, and so are
the other integrals of motion (speed and headway, respectively).

\begin{figure}[tbp]
\includegraphics[width=70mm,height=50mm]{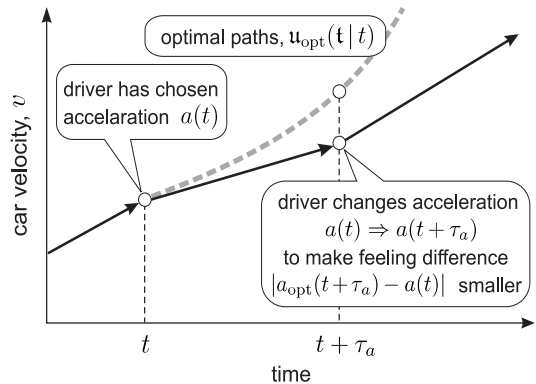}
\caption{The driver strategy of governing the car motion.\label{F1}}
\end{figure}

\section{Model description}

Assume that at a certain instant of time $t$ the driver has decided to
correct the car motion and chosen the acceleration $a(t)$
(Fig.~\ref{F1}). As discussed above, the optimal path
$\left\{\mathfrak{h}_{\text{opt}}(\mathfrak{t},t)\right\}$ of the
further motion ($\mathfrak{t}>t$) is too complex for her to compute
and to follow it. So, she regards the path
$\left\{\mathfrak{h}_{a}(\mathfrak{t},t):\mathfrak{a}(\mathfrak{t},t)=a(t)\right\}$
characterized by the constant acceleration as the optimal one.

A certain time interval $\tau_{a}$ later, the driver has to correct
the car motion again. This can be done by shifting the current
acceleration $a(t+\tau_{a})$ towards the desired optimal value
$a_{\text{opt}}(t+\tau_{a})=-\partial
_{\mathfrak{t}}^{2}\mathfrak{h}_{\text{opt}}(\mathfrak{t},t+\tau
_{a})|_{\mathfrak{t}=t+\tau_{a}}$ known to her approximately:
\begin{equation*}
a(t+\tau _{a})-a(t) = C\left( a_{\text{opt}}(t+\tau _{a})-a(t) \right) +
a_{\text{rnd}}(t+\tau _{a})\,,  \label{m.1}
\end{equation*}
where $C \lesssim 1$ is a constant about unity and the random term
$a_{\text{rnd}}(t+\tau _{a})$ allows for the uncertainty in the driver
evaluation of the optimal acceleration at the current time. Its mean
amplitude $a_{c}$ characterizes physiological properties of drivers
and can be considered constant. Thereby, $\left\langle
  a_{\text{rnd}}(t)\cdot a_{\text{rnd}}(t^{\prime })\right\rangle =
a_{c}^{2} \delta_{t,t'}$, where $\delta_{t,t'}$ is Kronecker's delta.

This discrete representation of the car motion correction is converted
to a continuous description based on stochastic differential
equations. Namely, the above discrete governing equation is reduced to
%
\begin{equation}
\frac{da}{dt}=-\frac{1}{\tau _{a}}\left( a-a_{\text{opt}}(h,v,V)\right)
+\eta \xi(t)\,.  \label{m.3}
\end{equation}
Here, $a_{\text{opt}}(h,v,V)$ is the optimal acceleration specified by
the current values of headway, car velocity, and leading car velocity.
The term $\xi(t)$ is white noise of unit amplitude which models the
uncertainty in the driver evaluation of the optimal motion.

The acceleration increment $\delta a$ caused by the random force
$\eta\xi(t)$ acting during the time $\tau_{a}$ is actually the random
component $a_{\text{rnd}}(t)$ entering the discrete governing
equation. Thus, it follows from the estimate $\langle(\delta
a)^{2}\rangle \sim \eta^2 \tau _{a}$ that
\begin{equation}
\eta = \frac{a_{c}}{\sqrt{\tau_{a}}}\,.  \label{m.4}
\end{equation}

The time scale $\tau_{a}$ of the driver control over the car motion
depends on the state $(h,v,V,a)$. Thus, the stochastic differential
equation~(\ref{m.3}) contains multiplicative noise. So its type with
respect to the corresponding Fokker-Planck equation has to be
specified. The adopted driving strategy (Fig.~\ref{F1}) implies that
all the characteristics of correcting the car motion are determined by
its state at the ``terminal'' point $t+\tau_a$ rather than at the
``initial'' point $t$. Therefore, it is reasonable for Eq.~(\ref{m.3})
to be of Klimontovich type or, according to the classification in
\cite{MH}, to describe a ``postpoint'' random process.

To complete the model, $a_{\text{opt}}(h,v,V)$ and $\tau_a(h,v,V,a)$
have to be specified. The simple ansatz
\begin{equation}
a_{\text{opt}}(h,v,V)=-\frac{1}{\tau }\left[ \left( v-V\right) -\frac{1}{\tau }%
g_{h}\left( h-h_{V}\right) \right]  \label{f.1}
\end{equation}
is well justified, at least, near the stationary state of the car
motion, $v=V$ and $h=h_{V}$. It should be noted that similar ideas
about $a_{\text{opt}}(h,v,V)$ and a dependence of $\tau_a$ on the
motion state had been discussed already in Ref.~\cite{Hel}. (See also
\cite{D3} for a discussion.)

Here, $\tau$ is the characteristic time of the velocity variations and
the constant $g_{h}\lesssim 1$. The limit $g_{h}\ll 1$ deserves
special attention because it is just the condition that a driver, at
first, prefers to eliminate the velocity difference $v-V$ between her
car and the car ahead and only then optimizes the headway. In this
case the optimal dynamics of car motion, i.e., the car dynamics
governed by the relation $a=a_{\text{opt}}(h,v,V)$ is a pure fading
relaxation towards the stationary state. Conversely, the model under
consideration predicts complex oscillations in the car motion. Note,
that the adopted assumption about the value of the coefficient $g_{h}$
can be justified by applying to the general principles of the car
motion~\cite{we2}.

If the car motion state is far from equilibrium the necessity for
correcting the velocity and headway distance is obvious. In this case
it is natural to suppose that the characteristic time interval
$\tau_{a}$ between sequential attempts to correct the car motion
should be comparable to $\tau$ which characterizes the velocity
variations, i.e., $\tau _{a}\sim \tau/g_v$. Here, $g_{v}\gtrsim 1$ is
an additional model parameter. When the car motion comes close to the
equilibrium and the inequality $|a_{\text{opt}}(h,v,V)| \lesssim a_c$
is fulfilled the uncertainty $a_{\text{rnd}}(t+\tau _{a})$ in
evaluating the optimal acceleration becomes significant. Under such
conditions there is no reason for the driver to affect the car motion
and she may not correct it at all. It means that the car motion
control is depressed and, correspondingly, the correction time
interval $\tau_a$ grows dramatically inside a domain $\mathbb{Q}_{u}$
of the phase plane $\left\{ h,v,V\right\}$ where the inequality
$|a_{\text{opt}}(h,v,V)| \lesssim a_c$ holds.

\begin{figure}[tbp]
\includegraphics[width=60mm]{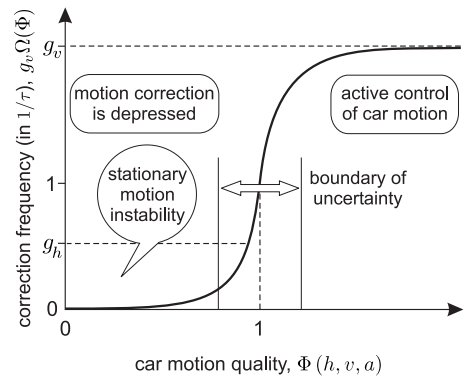}
\caption{The correction frequency $1/\tau _{a}$ of car motion control as
  function of the car motion quality $\Phi (h,v,V,a)$.\label{F2}}
\end{figure}

To compute the function $\tau_a(h,v,V,a)$, the boundary of the domain
$\mathbb{Q}_{u}$ has to be analyzed. Note, that the acceleration
itself enters the driver's perception of motion quality: without any
reason, a driver prefers not to accelerate at all. When the car motion
control is active the estimate $\dot{a}\sim a/\tau_{a}$ by virtue of
Eq.~(\ref{m.3}) can be adopted. So, the boundary of the domain
$\mathbb{Q}_{u}$ is specified by $a^2_{\text{opt}}(h,v,V) + \mu^2 a^2
\sim a^2_c$, where $\mu\sim 1$ is a certain coefficient about
unity. Assuming the variables $h$, $v$, $a$ to be independent of one
another inside $\mathbb{Q}_{u}$ and averaging the latter expression
over $\mathbb{Q}_{u}$ its boundary $\Phi (h,v,V,a) \sim 1$ can be
derived:
\begin{equation}
\Phi (h,v,V,a) = \frac{\left( v-V\right) ^{2}}{a_{c}^{2}\tau ^{2}}+g_{h}^{2}%
\frac{\left( h-h_{V}\right) ^{2}}{a_{c}^{2}\tau ^{4}}+\mu ^{2}\frac{a^{2}}{%
a_{c}^{2}}\,.  \label{f.5}
\end{equation}
If $\Phi (h,v,V,a) \ll 1$ the driver activity in correcting the car
motion is depressed completely. Otherwise, $\Phi (h,v,a) \gg 1$, the
driver controls the car motion actively. This is described by the
dependence of the correction time interval $\tau_{a}$ on the car
motion state,
\begin{equation}
\frac{1}{\tau _{a}}=g_{v}\Omega \left( \Phi (h,v,V,a)\right) \frac{1}{\tau }\,.
\label{f.6}
\end{equation}
The form of the function $\Omega \left( x\right) $ is illustrated in
Fig.~\ref{F2}. Equation~(\ref{m.3}) together with
expressions~(\ref{m.4}), (\ref{f.5}), and (\ref{f.6}) form the
proposed car following model with bounded rational drivers.

When $g_h > g_v\Omega(0)$ the stationary motion with $v=V$ and $h=h_V$
is unstable, leading to non-damped but bounded oscillations in the
headway and velocity of the following car. The particular form of the
function $\Omega \left( x\right)$ is of minor importance, it is only
necessary that its value inside $\mathbb{Q}_{u}$ to be small in
comparison with the ratio $g_h/g_v$. When analyzing the model
numerically the following ansatz
$$
    \Omega(x) = \exp[(x-1)/\Delta]/( \exp[(x-1)/\Delta] + 1 )
$$
is used, with the parameter $\Delta\sim 0.2$. Below, numerical results
will be presented that demonstrate the characteristic properties of
the developed model.

\begin{figure}[tbp]
\includegraphics[width=70mm]{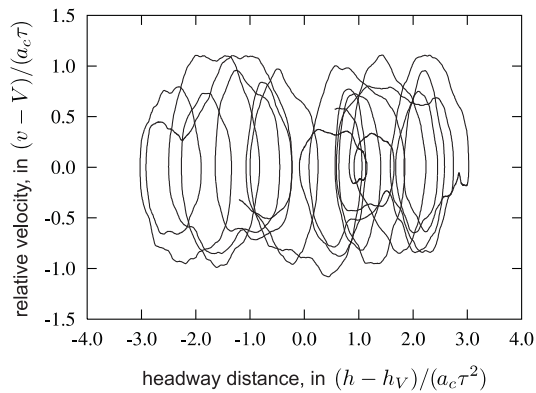}
\caption{Simulated car-following behavior. Integration of the stochastic
  differential equation has been performed with the algorithms
  described in \cite{SRK}. The parameters used are $g_v = 5$, $g_h =
  0.2$, $\mu = 1$, and $\Delta = 0.2$.\label{F3}}
\end{figure}

Figure~\ref{F3} displays an example of this dynamic in the $hv$-phase plane for
the dimensionless headway $x = (h-h_{V})/(a_{c}\tau ^{2})$ and the relative car
velocity $u =(v-V)/(a_{c}\tau)$. As seen in Fig.~\ref{F3}, the behavior of this
model is qualitatively similar to the empirical data in Fig.~\ref{F0}.

Preliminary results have shown that, first, the quasi-period of these
oscillations in the car velocity is equal to $\tau$ times a numerical factor
(about ten) depending weakly on the model parameters. For $\tau\sim 1$~s this
period is similar to the observed quasi-period. Second, the amplitude of
velocity oscillations does not change substantially as the model parameters
vary and is about $a_{c}\tau$. By visually comparing Figs.~\ref{F0}, \ref{F3}
the estimate $a_c\sim 0.3$~m/s$^2$ is obtained. It should be noted that the
amplitude of the acceleration oscillations exceeds $a_c$ by a numerical factor
about three. Third, the amplitude of the headway super-oscillations, in
contrast, depends essentially on the parameter $g_v$, enabling one to fix this
parameter based on experimental data.

\section{Summary}

A model regarding the bounded rational behavior of car drivers has
been supposed in this contribution. It takes into account that
drivers, although having detailed ideas about their preferred driving
strategy, are not able to control this driving strategy sufficiently
precisely. Namely, drivers introduce three main sources of error into
the optimal driving strategy: instead of keeping track of the changes
in acceleration they simply choose a constant one, that additionally
is not the optimal one but blurred by noise. This noise models the
inability of drivers to evaluate exactly the very complex integrations
leading to an optimal driving strategy. Therefore, the need to correct
the motion from time to time arises, with the correction time
intervals distributed randomly but inversely proportional to the
deviation from the desired optimal acceleration.

It is shown, that these ideas can be captured in a simple model for
the car-following dynamics, however at the cost of introducing
a non-Newtonian term, the jerk (change in acceleration). The benefit
of doing so is that the resulting model has smooth trajectories in
headway, velocity and acceleration but still being a stochastic one.
This discerns the approach proposed here from almost all models of
car-following introduced so far.

Although the trajectories generated by this model have some similarities with
real car-following data, the approach proposed here still needs thorough
testing with empirical data. This will be done in the near future and will be
reported soon.

These investigations were supported in part by RFBR
Grants~01-01-00389, 00439 and INTAS Grant~00-0847.


\begin{thebibliography}{99}

\bibitem{Chowdhury}
D.~Chowdhury, L.~Santen, and A.~Schadschneider,  Phys. Rep. \textbf{329},
199 (2000).

\bibitem{Helbing}
D.~Helbing, Rev. Mod. Phys. \textbf{73}, 1067 (2001).

\bibitem{daganzo-critique-sync}
C.~F.~Daganzo, M.~J.~Cassidy and R.~L.~Bertini, Transp.~Res.~A {\bf 33}, 365
(1999).

\bibitem{Kerner} B.~S.~Kerner and S.~L.~Klenov, J. Phys. A {\bf 35},
  L31 (2002).

\bibitem{BL} W.~Knospe, L.~Santen, A.~Schadschneider, and
  M.~Schreckenberg, J. Phys. A {\bf 33}, L477 (2000).

\bibitem{Tomer} E.~Tomer, L.~Safonov, and S.~Havlin, Phys.~Rev.~Lett.,
  {\bf 84}, 382 (2000).

\bibitem{83}
A.~Reuschel, \"{O}sterr. Ingen.-Archiv \textbf{4}, 193 (1950).

\bibitem{84}
L.~A.~Pipes, J. Appl. Phys. \textbf{24}, 274 (1953).

\bibitem{B1}
M.~Bando, K.~Hasebe, A.~Nakayama, A.~Shibata, and Y.~Sugiyama, Phys. Rev. E
{\bf 51}, 1035 (1995); Jpn. J. Ind. Appl. Math. {\bf 11}, 202 (1994)

\bibitem{B2}
M.~Bando, K.~Hasebe, K.~Nakanishi, A.~Nakayama, A.~Shibata, and Y.~Sugiyama, J.
Physique I {\bf 5}, 1389 (1995).

\bibitem{Hel}
W. Helly, in: \textit{Proceedings of the Symposium on Theory of Traffic Flow},
Research Laboratories, General Motors (Elsevier, New York, 1959), p.~207.

\bibitem{Fritz}
H.~T.~Fritzsche, Transp. Eng. Contr. \textbf{5}, 317 (1994).

\bibitem{Xing}
J. Xing, in: \textit{Proceedings of the Second Word Congress on ATT}, Yokohama,
November, 1739 (1995).

\bibitem{skPRE}
S.~Krau{\ss}, P.~Wagner, and Ch.~Gawron, Phys. Rev. E {\bf 55}, 5597 (1997).

\bibitem{H1}
D.~Helbing and B.~Tilch, Phys. Rev. E \textbf{58}, 133 (1998).

\bibitem{H2}
M.~Treiber, M.~Hennecke, and D.~Helbing, Phys. Rev. E \textbf{62}, 1805 (2000).

\bibitem{D1}
R.~W.~Rothery, \textit{Car Following Models}, in: \textit{Traffic Flow Theory},
N.~Gartner, C.~J.~Messer, and A.~K.~Rathi (eds.) (Transportation Research
Board, Special Report 165, 1992), Chap.~4.

\bibitem{D3}
M. Brackstone and M. McDonald, Trans. Res. F \textbf{2}, 181 (1999).

\bibitem{D2}
M.~Bando, K.~Hasebe, K.~Nakanishi, and A.~Nakayama, Phys. Rev. E, \textbf{58},
5429 (1998).

\bibitem{we1}
I.~Lubashevsky, S.~Kalenkov, and R.~Mahnke, Phys. Rev. E \textbf{65}, 036140
(2002).

\bibitem{q}
T.~H.~Chang and I-S.~Lai, Transp. Res. C \textbf{6}, 333 (1997).

\bibitem{we2}
I.~Lubashevsky, P.~Wagner, and R.~Mahnke, to be published.

\bibitem{BR1} H.~A.~Simon, \textit{Theories of Bounded Rationality},
in: \textit{Decision and Organization} C.~B.~McGuire and R.~Radner (eds.)
(North-Holland, Amsterdam, 1972) (Chap.~8).

\bibitem{we3}
I.~Lubashevsky, R.~Mahnke, P.~Wagner, and S.~Kalenkov, Phys. Rev. E
\textbf{66}, 016117 (2002).

\bibitem{MH}
T.~Morita and H.~Hara, Physica A \textbf{101}, 283 (1980); \textbf{125}, 607
(1984).

\bibitem{SRK}
K.~Burrage and P.~M.~Burrage, Appl. Numer. Math. \textbf{22}, 81 (1996).

\end{thebibliography}
\end{document}